\documentclass[prb,reprint]{revtex4-1}
\usepackage{graphicx}
\usepackage{amsmath,bm}
\usepackage{color}
\usepackage{multirow}

\begin{document}

\newcommand{\ztc}{C$_{36}$H$_9$}
\newcommand{\substitute}[2]{\textcolor{red}{{#1}} \textcolor{blue}{{#2}}}

\title{
Magneto-orbital effect without spin-orbit interactions \\ 
--- noncentrosymmetric zeolite-templated carbon structure
}

\author{Takashi Koretsune$^1$, Ryotaro Arita$^{2,3,4}$ and Hideo Aoki$^5$}
\affiliation{
        $^1$Department of Physics, Tokyo Institute of Technology, Oh-okayama, Tokyo 152-8551, Japan \\
	$^2$Department of Applied Physics, University of Tokyo, Hongo, Tokyo 113-8656, Japan \\
	$^3$Japan Science and Technology Agency (JST), CREST, Honcho, Kawaguchi, Saitama 332-0012, Japan \\
	$^4$Japan Science and Technology Agency (JST), PRESTO, Kawaguchi, Saitama 332-0012, Japan \\
	$^5$Department of Physics, University of Tokyo, Hongo, Tokyo 113-0033, Japan
}
\date{\today}

\begin{abstract}
A peculiar manifestation of orbital angular momentum is proposed for 
a zeolite-templated carbon system, \ztc.   The structure, being a network of nanoflakes in the shape of a ``pinwheel", lacks  inversion symmetry.  
While the unit cell is large, 
the electronic structure obtained with a first-principles density functional theory and captured as an effective tight-binding model in terms of maximally-localized Wannier functions, exhibits an unusual 
feature that the valence band top comes from two 
{\it chiral} states having orbital magnetic momenta of $\pm 1$.   The 
noncentrosymmetric lattice structure then makes the band dispersion asymmetric, as reminiscent of, but totally different 
from, spin-orbit systems.  The unusual feature is predicted to 
imply a current-induced orbital magnetism when holes are doped.  
\end{abstract}


\maketitle

%{\it Introduction---}
\section{Introduction}
Functional materials, as exemplified by 
magnetoelectronic materials, ferroelectrics, thermoelectrics, or unconventional superconductors, usually rely essentially on strong Coulomb correlations or spin-orbit coupling in the system. That is why magnetic elements or heavy elements play a crucial role there, as in the unconventional superconductivity involving transition-metal compounds\cite{SCreview} and topological insulators involving 
heavy elements\cite{TIreview1,TIreview2}.  All the more acutely, 
we can pose a converse question: can we explore  possibilities to 
realize such unique physical properties in {\it light-element} 
materials, by designing unconventional structures?   

Among light elements, carbon is unique and promising.  Its covalent nature enables the atoms to have a variety of structures, ranging from three-dimensional diamond down to zero-dimensional fullerene, and the electronic structure dramatically changes accordingly. The scale of the band width goes from $\sim 10$ eV  in graphene\cite{Neto} down 
to $<1$ eV in solid fullerene\cite{Saito,Capone}. Carbon nanotube becomes a metal or semiconductor depending on its chirality\cite{Hamada}. Indeed, there is 
a body of theoretical and experimental investigations on functional materials made from carbon.  
For instance, Shima and one of the present authors theoretically proposed that long-period (i.e., antidot-array) graphene structures accommodate 
a unique platform for ferromagnetism as well as a manipulation between metallic and semiconducting band structures\cite{shimaaoki}.  

Now, an even more intriguing avenue that may harbor rich possibilities in electric properties is,  in our view, three-dimensional 
periodic networks of  carbon atoms.  A recent remarkable example due to Kyotani and coworkers is the zeolite-templated carbon (ZTC), which is a unique class of carbon networks fabricated by employing a zeolite framework as a template  (where zeolite is taken away after fabrication)\cite{kyotani1997,nishihara2009}.    A fascination with zeolite is its 
versatile (almost 200 known) structures\cite{zeoliteassociation}, so 
various ZTCs are expected to be possible.   Another fascination 
is zeolites' open structures with large cages arranged with long periods.  
In fact, regularly arranged alkali clusters in zeolite exhibit a variety of unusual physical properties,  including ferromagnetism\cite{Nozue,Nozue2} 
despite the system comprising only nonmagnetic elements.
An important theoretical 
picture that has emerged from the system is that the electronic structure of the alkali-metal loaded zeolites can be captured in the ``superatom" model\cite{AritaZeolite,Nohara1,Nohara2,SOD}, where the whole electronic band structure can be described surprisingly accurately as a tight-binding model of Wannier orbitals, which reside on each zeolite cage with a linear dimension $\sim 10$ \AA\ but can have symmetries such as s and p.   A message suggested by this is we may design versatile functional systems by controlling the arrangement of superatoms and/or the nature of each superatom.   

In the present work we go one step further in 
seeking a possibility of realizing unusual electronic structures from 
unusual geometries.  We shall propose that 
a zeolite-templated carbon can indeed harbor, when its crystal structure 
is noncentrosymmetric,  
an unusual,  asymmetric band dispersion arising from 
``chiral" superatoms having orbital magnetic momenta of $\pm 1$.  
So in this paper, we first obtain, with an {\it ab initio} density-functional calculation, the electronic structure of ZTC, for which the zeolite Y is used as the template. While this system has been experimentally studied intensively due to its high hydrogen storage ability\cite{kyotani1997,nishihara2009}, theoretical studies have been quite limited, since the crystal structure is extremely complicated and has not been determined experimentally. Recently, Kyotani {\it et al.} have constructed several possible atomic structure models\cite{nishihara2009}.  There, one possibility is a kind of network of carbon nanotubes\cite{suzuki}.   
Another possibility, on which we focus here, is a network of 
nanographenes in a form of buckybowl (as opposed to buckyball) in a three-dimensionally regular structure.   Here we take the simplest model, in which the building block is \ztc\ nanographene with hydrogen termination (called \ztc\ patch hereafter) (Fig.\ 1(a)). We shall show that, while there are as many as 
24 valence bands due to the complicated structure, it is possible to 
construct a simplified tight-binding model in terms of the maximally-localized Wannier functions for these bands, where the essence is a surprisingly simple three-orbital model on the diamond lattice with eight lattice points per unit cell.  The three orbitals, $\psi_1, \psi_2,$ and $\psi_3$, on each \ztc\ patch can be seen, respectively, as $p_x$, $p_y$ and $p_z$ orbitals of a superatom with the orbital angular momenta $l_z=-1,0,1$, 
although the symmetry of the three orbitals is lower than that of 
the atomic $p$ orbitals.

Now, what is unusual is the following: the present crystal structure may be viewed as a 
network of (three-winged) ``pinwheels" (consisting of four \ztc\ patches; Fig.\ 1(b)), where each pinwheel has a threefold axis but lacks the inversion symmetry 
(hence its name).  This deprives the crystal structure of the inversion symmetry as well, which makes valence top made from {\it chiral states}, 
which are symbolically 
$\psi^{\pm}_{\rm E} \propto (p_x + e^{\pm 2 \pi i/3} p_y + e^{\pm 4 \pi i/3} p_z)$, which belongs to the irreducible representation of E (see below). 
The two chiral bands are shown to have different band dispersions, which reminds us of spin-orbit split bands in GaAs but the cause of the splitting in the former is totally different.  
We finally predict that we can induce an electric-field-induced orbital magnetism if we dope carriers into the system and apply an external electric field. 
Although the asymmetric band dispersion has nothing to do with spin-orbit interaction, the unusual feature implies, in its own right, that a current-induced orbital magnetism is expected by hole doping.

%{\it Method---}
\section{Method}
To obtain the electronic properties, 
we have performed a first-principles calculation\cite{espresso}
within the local-density approximation (LDA)\cite{ceperley1980,perdew1981}.  
Ultrasoft pseudopotentials are used to describe the electron-ion interaction\cite{vanderbilt1990}.
The valence wave functions and charge densities are expanded by a plane-wave basis with cutoff energies of 30 Ry and 150 Ry, respectively. 
Integration over Brillouin zone is carried out with $2 \times 2 \times 2$ $k$ points.
We have employed the wannier90 code\cite{marzari1997} to compute the maximally-localized Wannier orbitals.

\section{Results}
%{\it Results---}
%---------------------------------------------------------------------------------------------------------------------------
\begin{figure}
	\includegraphics[scale=0.45]{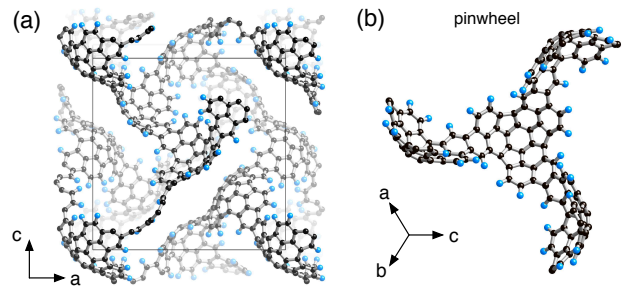}
\caption{
	(Color online)
  (a) Crystal structure, theoretically optimized here, of 
the zeolite-templated carbon.  
  Solid line represents a unit cell, which contains eight hydrocarbon \ztc\ patches (with 288 carbon (darker spheres) and 72 hydrogen atoms (lighter)) with different orientations.
  (b) A \ztc\ patch with its three nearest-neighbor ones, which form a three-winged ``pinwheel" with a threefold axis.
}
\label{fig:structure}
\end{figure}
%---------------------------------------------------------------------------------------------------------------------------
Figure \ref{fig:structure} shows the geometry of \ztc.
The structure is theoretically optimized with the experimental lattice constant of $a = $ 24.07 \AA.  
Since the zeolite Y 
template has a diamond-type structure,
the resulting \ztc\ is diamond-like as well.  
However, unit cell of the whole structure is simple-cubic rather than face-centered cubic, 
and contains eight hydrocarbon \ztc\ patches, whose normal directions are all different, $i.e.$, ($\pm 1$,$\pm 1$,$\pm 1$).
Each \ztc\ patch, connected to three nearest-neighbor ones as shown in Fig.\ \ref{fig:structure}(b),  form a three-winged ``pinwheel" with a threefold axis.  
This is how the crystal structure has a high symmetry (space group: P4$_1$32), but lacks inversion symmetry.

%---------------------------------------------------------------------------------------------------------------------------
\begin{figure}
	\includegraphics[scale=0.32]{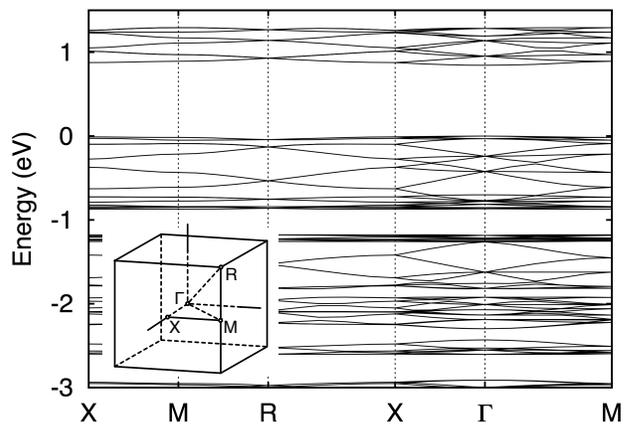}
\caption{
  LDA band structure of \ztc. Origin of energy is set to be the top of the valence band.  Inset shows the first Brillouin zone.
}
\label{fig:band}
\end{figure}
%---------------------------------------------------------------------------------------------------------------------------
The obtained electronic band structure of \ztc\ in Fig.\ \ref{fig:band} 
shows that \ztc\ is an insulator with a band gap of 0.84 eV.  
Since the three-dimensional network comprises nanoflakes, the 
band structure resembles those of molecular crystals where weakly coupled molecular orbitals result in a narrow band widths.  
The valence band top ($-1 < E < 0$ in eV) comprises 24 bands that are isolated from lower ones with an energy gap.
Interestingly, the top and bottom of this region are delineated by 
almost flat bands.
The band widths of these nearly flat bands are around 0.05 eV,
which is much smaller even when compared with 
that of t$_{1u}$-band in fullerene compounds ($\sim$ 0.3 - 0.5 eV).
Naively, one might expect magnetic or other instabilities with hole doping for these flat bands, but 
we have not found any magnetic orders such as ferromagnetism or antiferromagnetism within the local spin-density approximation (LSDA).

%---------------------------------------------------------------------------------------------------------------------------
\begin{figure*}
	\includegraphics[scale=0.65]{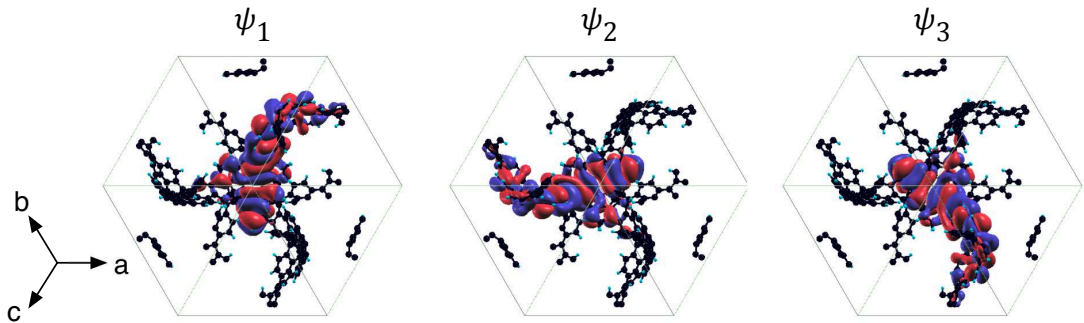}
\caption{
	(Color online)
  Maximally-localized Wannier functions, $\psi_1$, $\psi_2$, and $\psi_3$, for the 24 bands at the top of the valence band.
  The surfaces with different hues represent positive (negative) isosurfaces.
}
\label{fig:mlwfs}
\end{figure*}
%---------------------------------------------------------------------------------------------------------------------------
Let us then characterize these bands in terms of maximally-localized Wannier functions.  
Figures \ref{fig:mlwfs}(a)-(c) visualize the maximally-localized Wannier orbitals for the 24 bands.
On each ``pinwheel" reside three, equivalent Wannier orbitals that spread, respectively, along the three ``wings" (i.e., along the three next-nearest directions).  
These three orbitals, which we call $(\psi_1, \psi_2, \psi_3)$, conform to the threefold symmetry of the pinwheel, 
and, in the superatom language, roughly correspond to $(p_x, p_y, p_z)$ orbitals. Thus we may regard the system as a ``$p$-electron superatom system" on the diamond lattice.  However, crucial 
differences between the present superatom system and usual diamond 
are:  (i) there are no $s$-orbitals, and  
(ii) $p$-orbitals do not have full symmetries that the atomic $p$-orbitals have
and are symmetric only under a three-fold axis rotation.
This can be seen from the fact that each $p$-orbital has its center of wave function 
shifted from the center of the superatom (i.e., the center of the pinwheel) 
unlike the atomic $p$-orbitals.

%---------------------------------------------------------------------------------------------------------------------------
\begin{figure}
	\includegraphics[scale=0.65]{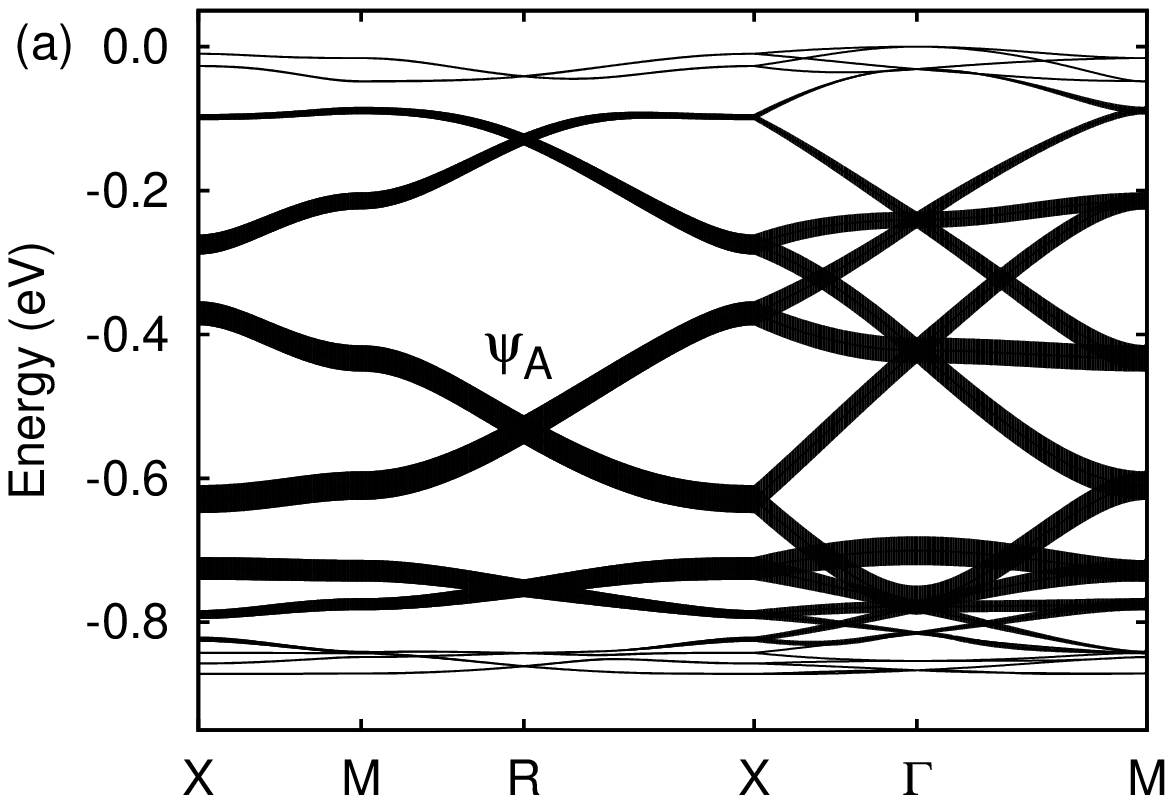}

	\includegraphics[scale=0.65]{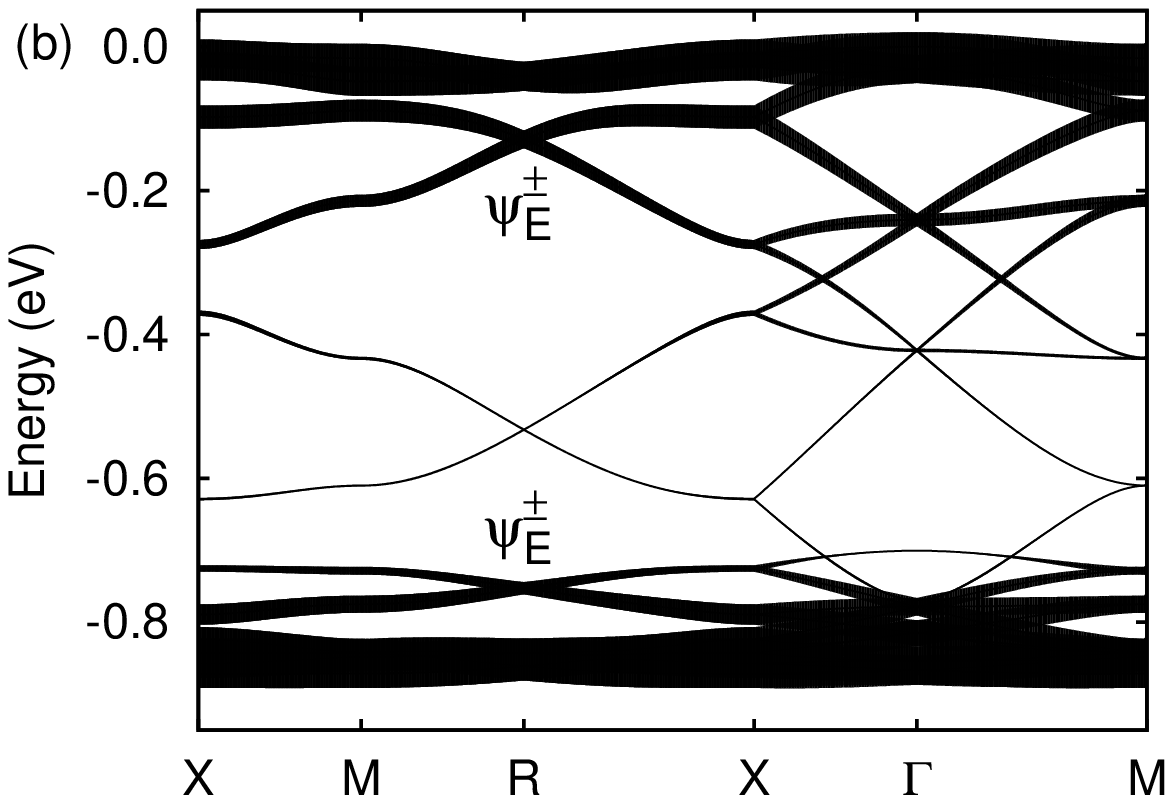}
\caption{
Structure of 24 topmost valence bands with the line width 
representing the weights of (a) $\psi_{\rm A}$ state, 
or (b) $\psi^{\pm}_{\rm E}$.
}
\label{fig:weight}
\end{figure}
%---------------------------------------------------------------------------------------------------------------------------
Due to the symmetry of the present $p$-orbitals, these valence bands should be characterized
by an A state and two E states given as
\begin{align}
	\psi_{\rm A} &= \frac{1}{\sqrt{3}} \left( \psi_1 + \psi_2 + \psi_3 \right),\\
	\psi_{\rm E}^{\pm} &= \frac{1}{\sqrt{3}} \left( \psi_1 + e^{\pm 2 \pi i/3} \psi_2 + e^{\pm 4 \pi i/3} \psi_3 \right),
\end{align}
where 
$\psi_{\rm A}$ ($l_z=0$ state) is the symmetric combination of the three Wannier functions, while
$\psi_{\rm E}^{\pm}$  ($l_z=\pm1$ states) have finite angular momenta associated with rotation around the threefold axis.  
With these states it is found that, the doubly-degenerate valence-top wavefunctions, for example out of the 24 bands, comprise entirely from $\psi_{\rm E}^{\pm}$ at the $\Gamma$ point from the symmetry analysis\cite{note_sym}.  
The calculated weights of $\psi_{\rm A}$ and $\psi_{\rm E}^{\pm}$ are displayed in Figs.\ \ref{fig:weight}(a) and (b), respectively.
Interestingly, the nearly flat bands along the top and bottom of
the 24 bands originate from the finite-angular-momentum $\psi^{\pm}_{\rm E}$ states, 
while the dispersive bands from $\psi_{\rm A}$.

\begin{figure}
	\includegraphics[scale=0.4]{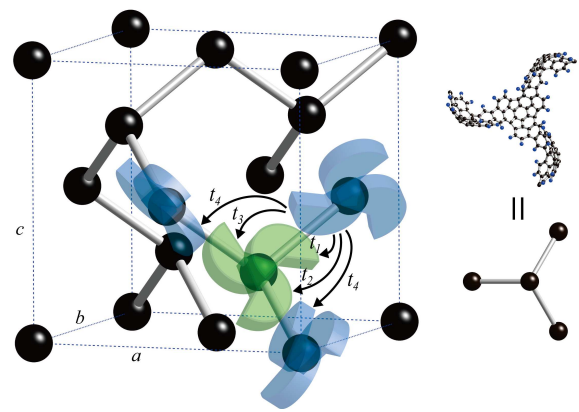}
\caption{
	(Color online)
The \ztc\ represented by the obtained tight-binding model.
Each black sphere represents a \ztc\ patch, while wings of a pinwheel illustrate Wannier orbitals.  The transfer integrals are: 
  $t_1 = -0.255$ eV, $t_2 = 0.116$ eV, $t_3 = 0.022$ eV, and $t_4 = -0.092$ eV.
}
\label{fig:tightbinding}
\end{figure}
%---------------------------------------------------------------------------------------------------------------------------

Figure \ref{fig:tightbinding} illustrates the tight-binding model for the 24 bands obtained in terms of the maximally-localized Wannier orbits 
(wings of a pinwheel in the figure).  
Since the maximally-localized Wannier functions are basically 
localized on each patch, it is natural that the $3\times 8 =24$ bands are 
all accurately described by them.  
There are three transfer integrals, $t_1= -0.255$ eV, $t_2 = 0.116$ eV, and $t_3 = 0.022$ eV, between nearest-neighbor patches, 
while the largest transfer integral between second-neighbor patches 
is $t_4 = -0.092$ eV. These exhaust the transfer integrals whose 
magnitudes exceed $0.01$ eV.  Although the unit cell of the ZTC is huge and complicated (containing more than 300 atoms), these four transfer integrals 
suffice to reproduce the dispersions of the 24 valence bands around the Fermi level.

%{\it ``Orbital-Rashba" effect ---} 
%{\it $l_z$-separated dispersion and current-induced orbital magnetism ---} 
\section{Separated dispersion and current-induced orbital magnetism}
\begin{figure}
	\includegraphics[scale=0.65]{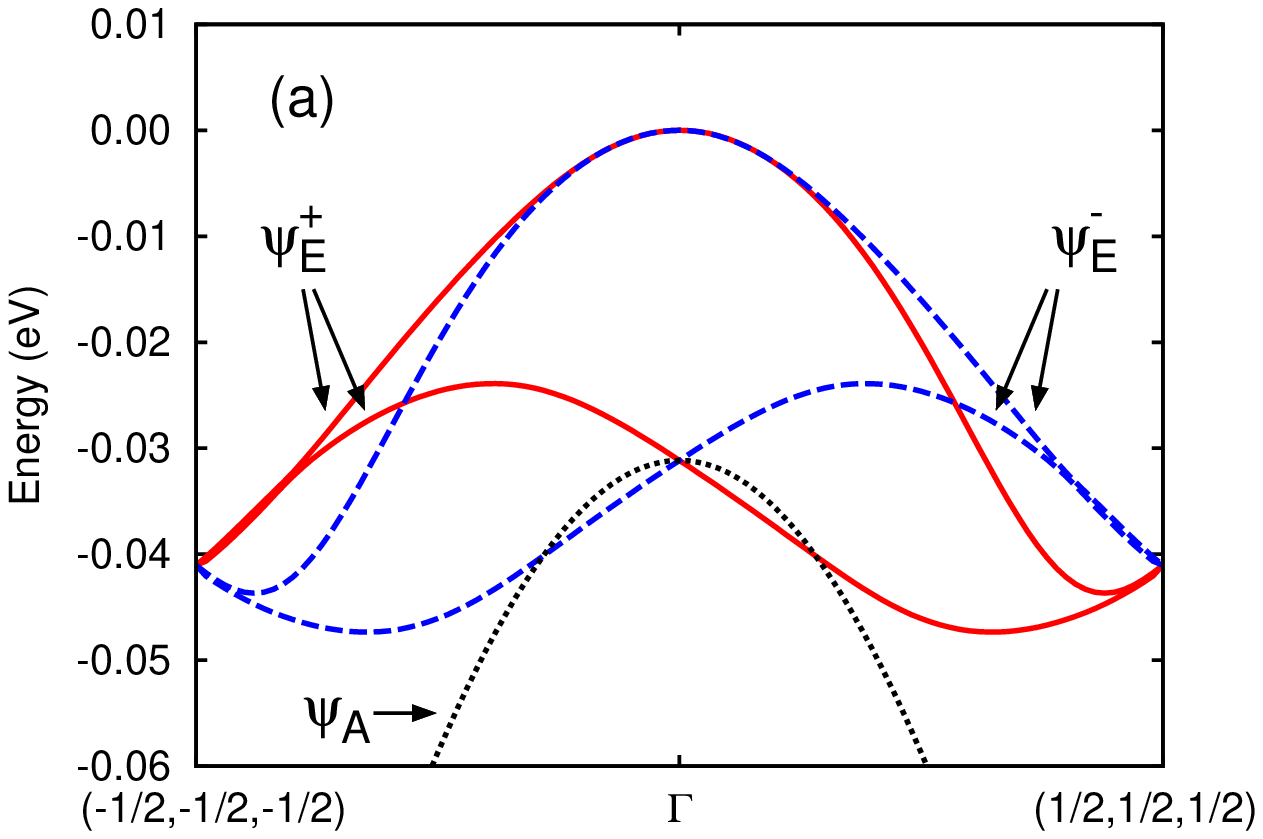}

	\includegraphics[scale=0.65]{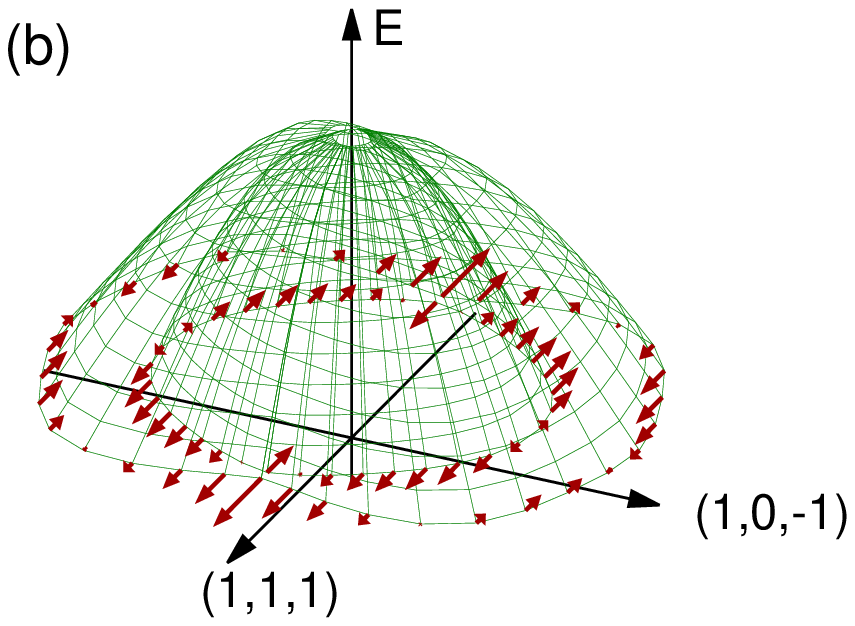}

	\includegraphics[scale=0.65]{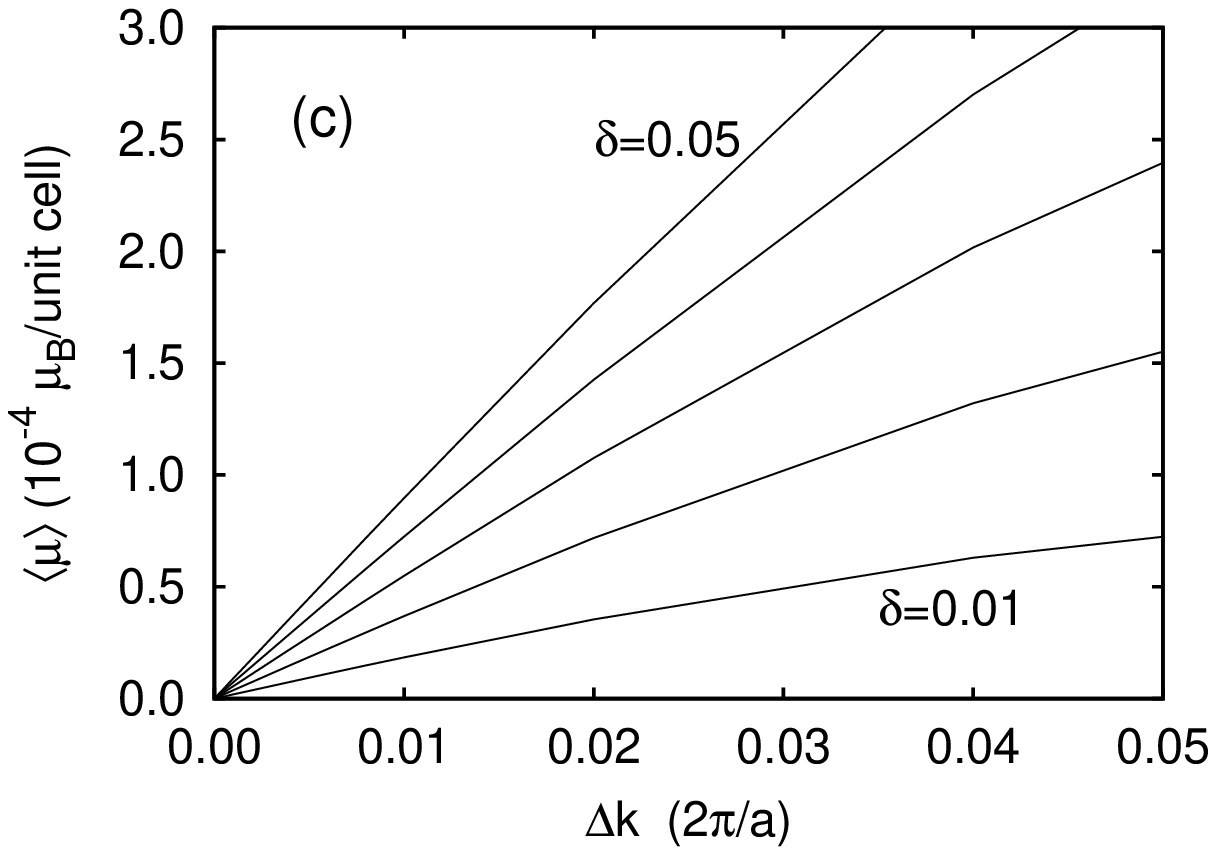}
\caption{
	(Color online)
(a) A blow-up of Fig.\ref{fig:weight}, where  bands originating from 
 $\psi_{\rm A}$ (black dotted line), 
$\psi^{+}_{\rm E}$ (red solid), 
and $\psi^{-}_{\rm E}$  (blue dashed) on a 
patch normal to (1,1,1) are displayed, respectively, 
along $\Gamma$ - R line.  
(b) Two topmost valence bands plotted on a two-dimensional $k-$space.  
The  two bands anticross with each other along the (1,1,1) direction, and 
arrows represent the orbital moment, $l_z$.
(c) The total orbital magnetic moment per unit cell as a function of the electric field with the number of holes, $\delta = 0.01, 0.02, 0.03, 0.04$ and $0.05$ (from bottom to top).  
The applied electric field is here represented by $\Delta k$, which 
is the crystal momentum acquired by an electron due to the field in (1,1,1) direction.
}
\label{fig:weight2}
\end{figure}

Now we come to the key result in the present work in 
Fig.\ref{fig:weight2}, which depicts the structure of the 
topmost five valence bands, with the main character of 
each band indicated in terms of $\psi_{\rm A}$, $\psi^{+}_{\rm E}$, and $\psi^{-}_{\rm E}$.  
Remarkably, the two topmost bands are {\it asymmetric} with respect to $\Gamma$ point.  They arise from the absence  of inversion symmetry in the crystal structure, and transform with each other by the time-reversal.  

The feature may seem reminiscent of the noncentrosymmetric systems with strong spin-orbit coupling, but the cause of the splitting is totally 
different here.   A conventional spin-orbit model is Rashba's\cite{rashba1984}: 
in a two-dimensional electron gas with a strong spin-orbit coupling, 
the spin is no longer a good quantum number, but the bands are still degenerated due to the Kramers degeneracy.  
If we apply an electric field perpendicular to the plane, the degeneracy is lifted, and a vortex-like spin texture is generated.  Even in the absence of electric fields, various spin textures can emerge if the system is inversion asymmetric\cite{Dresselhaus,Oguchi}. In fact, the original proposal by Rashba was made for noncentrosymmetric wurtzite semiconductors\cite{Rashbaorg,Ishizaka, Saeed}.  Such systems harbor a possibility of manipulation of the spin degree of freedom by applying an external electric field\cite{Sinova, Murakami}.  
In materials belonging to the gyrotropic point group, charge current is generically accompanied by a nonzero spin polarization\cite{Ivchenko}.

By contrast, in the present ZTC 
we can expect a {\it current-induced orbital magnetism}: If we dope holes 
to extract the peculiar property of the valence top
and apply an external electric field, 
the charge carriers drift in the direction of the applied field, so that 
the population of the $l_z=1$ states will exceed that of $l_z=-1$ as in the spin-orbit case\cite{edelstein1990}, 
since the band structure of the two orbital-magnetic states is asymmetric around the $\Gamma$ point as shown in Fig.\ \ref{fig:weight2}\cite{park2012}.
For example, if we dope 0.05 holes per unit cell and apply an external electric field in (1,1,1) direction so that the electrons acquire the quasimomentum $\Delta k = 2 \times 10^{-2}$ in the unit of $2\pi/a$, 
the resulting orbital magnetic moments, in units of 
$10^{-4} \mu_{\rm B}$ per unit cell, are
$\langle \mu_{(1,1,1)} \rangle = 0.47$, 
$\langle \mu_{(-1,-1,-1)} \rangle = -0.50$, 
$\langle \mu_{(1,1,-1)} \rangle = \langle \mu_{(1,-1,1)} \rangle = \langle \mu_{(-1,1,1)} \rangle = 0.40$, and
$\langle \mu_{(-1,-1,1)} \rangle = \langle \mu_{(-1,1,-1)} \rangle = \langle \mu_{(1,-1,-1)} \rangle = -0.39$, 
where $\langle \mu_{(a,b,c)} \rangle$ denotes the orbital magnetic moment contributed from \ztc\ patch whose normal direction is $(a,b,c)$.  
Since the quantization axis differs from one patch to another, these moments do not cancel with each other, and the total orbital magnetic moment in (1,1,1) direction is 
$\langle \mu \rangle = \sum_{a,b,c=\pm 1} \langle \mu_{(a,b,c)} \rangle (a+b+c)/3 = 1.8 \times 10^{-4} \mu_{\rm B}$.  When the electric field 
and/or the number of holes are 
varied, the magnetic moment increases 
roughly proportional with the field and with 
the number of holes as shown in Fig.\ \ref{fig:weight2}(c).

The current-induced orbital magnetism in the present mechanism 
exploits the fact that electrons are confined on the curved surface of the \ztc\ patch. 
Namely, in an electric field along, say, (1,1,1), 
electrons on a pinwheel $\perp (1,1,1)$ cannot move along that direction, but edge currents are generated instead. Due to inversion asymmetry, the right-handed and left-handed edge currents do not cancel with each other, which will induce an orbital magnetic moment. 
Thanks to the small width of the relevant bands, a relatively small electric field should suffice to induce such an orbital magnetism.
This contrasts with the usual exploitation of magnetoelectric effects in solids for spintronics\cite{kato2004,sih2005}, where appreciable spin-orbit interaction is required to couple the motion of charge with the spin degree of freedom.  
The present mechanism for the current-induced orbital magnetism requires no spin-orbit coupling, so that we may expect that this mechanism will open a new avenue for light-element functional materials.
Further systematic studies on electronic structures of ZTC systems is an interesting future problem.

%---------------------------------------------------------------------------------------------------------------------------
%---------------------------------------------------------------------------------------------------------------------------

%{\it Conclusion---}
\section{Conclusion} 
We have found that the zeolite-templated carbon system, \ztc, has a unique electronic structure around the top of the valence band, where orbital character is {\it asymmetric} around $\Gamma$ point due to a combined effect
of the chiral nature in the orbital character and the lack of inversion symmetry.  As a consequence of this characteristic band structure, a {\it current-induced orbital magnetism} is predected 
when holes are doped.  
An orbital magnetic texture, which resembles spin textures in relativistic noncentrosymmetric systems, is then expected to appear despite the absence of spin-orbit coupling, and can be understood in terms of the simple tight-binding model.
 
{\it Acknowledgments---}
The authors would like to thank Takashi Kyotani and Shuichi Murakami 
for valuable discussions, and Sergey Ganichev for 
illuminating the group theoretical aspects.  
Numerical calculations were performed on TSUBAME Grid Cluster at Global Scientific Information and Computing Center of the Tokyo Institute of Technology. This work was partially supported by Grant-in-Aid for Scientific Research from MEXT Japan under contract numbers 19051016 (TK, RA, and HA), JST-PRESTO, Funding Program for World-Leading Innovative R\&D on Science and Technology (FIRST program) on ``Quantum Science on Strong Correlation'', 
and the Computational Materials Science Initiative (CMSI), Japan (RA).

%\bibliographystyle{apsrev4-1}
%\bibliography{manuscript}

\begin{thebibliography}{37}
\bibitem [{not()}]{note_sym}%
  \BibitemOpen
  \href@noop {} {}\bibinfo {note} {At the $\Gamma$ point, irreducible
  representations for wavefunctions should be any one of A$_1$, A$_2$, E,
  T$_1$, and T$_2$ and the wavefunction belonging to E (A$_1$ or A$_2$)
  representation consists only of $\psi_{\rm E}^{\pm}$ ($\psi_{\rm A}$) for
  each of eight patches.}\BibitemShut {Stop}%
\end{thebibliography}

%merlin.mbs apsrev4-1.bst 2010-07-25 4.21a (PWD, AO, DPC) hacked
%Control: key (0)
%Control: author (72) initials jnrlst
%Control: editor formatted (1) identically to author
%Control: production of article title (-1) disabled
%Control: page (0) single
%Control: year (1) truncated
%Control: production of eprint (0) enabled
%

\end{document}